# Giant electron-phonon anomaly in doped La$_2$CuO$_4$ and other cuprates

D. Reznik

Forschungszentrum Karlsruhe, Institut für Festkörperphysik, D-76121 Karlsruhe, Germany

**Since conventional superconductivity is mediated by phonons, their role in the mechanism of high temperature superconductivity has been considered very early after the discovery of the cuprates. The initial consensus was that phonons could not produce transition temperatures near 100K, and the main direction of research focused on nonphononic mechanisms. Subsequent work last reviewed by L. Pintschovius in 2005 showed that electron-phonon coupling in the cuprates is surprisingly strong for some phonons and its role is controversial. [1] Experiments performed since then identified anomalous behavior of certain Cu-O bond-stretching phonons in cuprates as an important phenomenon that is somehow related to the mechanism of superconductivity. A particularly big advance was made in the study of doped La$_2$CuO$_4$. This work is reviewed here.**

**Introduction**

In most cases superconductivity is mediated by phonons and the transition temperature T$_c$ scales with the strength of electron-phonon coupling. In conventional superconductors with relatively high T$_c$s (>10K), it is often possible to identify specific modes exceptionally strongly coupled to electrons. Such phonons are softer and broader than expected from lattice-dynamical models. Knowing what these modes are makes it possible to influence T$_c$ by variations in chemical compositions, pressure or other parameters. For a long time inelastic neutron scattering (INS) has been the only technique allowing their direct measurements throughout the Brillouin zone. In recent years high resolution inelastic x-ray scattering (IXS) emerged as an alternative technique with some advantages and disadvantages vs. INS.

Shortly after the discovery of the high T$_c$ cuprates, extensive effort focused on looking for effects of strong electron-phonon coupling. This task proved a lot more challenging than in conventional superconductors, because due to the large unit cell, the cuprates have many phonon branches. In addition, most soft phonons in conventional superconductors appear in acoustic branches at low energies (with a notable exception of MgB$_2$), whereas we now know that such phonons in the cuprates are at much higher energies and belong to optic branches. Furthermore, unlike in conventional superconductors, standard theory could not predict soft phonon behavior in the cuprates, so the identification of these modes had to be done by "blind search". Another obstacle was that mapping phonon dispersions by neutron scattering requires large single crystals, which became available years after the initial discoveries of some cuprates and are still not available for others. The field received an important boost after the development of high resolution inelastic x-ray scattering (IXS). [2] It allows measurements of very small samples greatly increasing the range of materials that can be studied. Compared to neutron scattering it has a superior wavevector resolution and a superior energy resolution at large energy transfers. Some of its main drawbacks are scarcity of beamtime, lower resolution at low energies than INS, Lorentzian resolution function (vs. Gaussian for INS), and weak intensities of high energy phonons in some materials. In the study of cuprates INS and IXS are used as complementary techniques. In recent years they provided important new insights reviewed here.

In order to establish proper context, I will begin by reviewing some old and new results on conventional systems. Most of the article will focus on the giant bond-stretching phonon anomaly in $La_2CuO_4$-based family of the high $T_c$ cuprates where most progress has been made in the last few years. Finally, I will present recent results for other cuprates to demonstrate that this phonon effect is not limited to doped $La_2CuO_4$.

## 1. Electron-phonon coupling in conventional metals
### 1.1 General Considerations

Electron-phonon coupling typically enters theoretical frameworks as a cross section (matrix element) for scattering of a single valence electron by a single phonon. Unfortunately there is no experimental technique allowing direct measurements of this scattering cross section. The closest one can do is to measure either spectral functions of specific phonons by INS and IXS or to measure electronic Green's functions by angle resolved photoemission (ARPES). The phonon spectral functions are influenced by scattering of a specific phonon by all electrons (electronic contribution to the phonon self-energy), whereas the electronic Green's function includes scattering of a specific electronic state by all phonons (phonon contribution to the electronic self-energy).

This review focuses on INS and IXS measurements of phonon spectral functions. Phonons couple to electronic polarizability, i.e. the two-particle electronic response. Usually the electron-phonon coupling is weak and there are no singularities in the electronic response close to the phonon energy. In this case the phonon spectral function is that of a damped harmonic oscillator. In the limit of weak damping it is close to a Lorentzian centered at the phonon frequency whose linewidth is proportional to the inverse phonon lifetime.

Electron-phonon scattering renormalizes the phonon frequency and reduces the lifetime (real/imaginary parts of the phonon self-energy). The amount of this renormalization depends on the phase-space available for the scattering of electrons by the phonon as well as on the cross section (matrix element) for each of these electron-phonon scattering processes.

Isolating this electronic contribution requires knowledge of frequencies and linewidths in the absence of electron-phonon coupling, since anharmonicity, inhomogeneity, and impurities may have the same effect as electron-phonon coupling. Determining these accurately is typically a challenge. In fact even defining the real part of the phonon self-energy, is not at all straightforward (see sec. 1.3). Thus very precise measurements of phonon frequencies and linewidths do not, in a general case, provide direct information on electron-phonon coupling. There are, however, special cases, where one can be fairly certain of the role of the electronic contribution to the phonon self-energy.

## 1.2 Kohn anomalies

The simplest way to model phonons is by balls-and-springs models with the atomic nuclei serving as balls and the Coulomb forces screened by the electrons as springs. Shell models are more sophisticated modifications of this approach. Including only short-range interactions gives smooth phonon dispersions. Such models can be fit to the experimental dispersions that do not contain any sharp dips. In metals with Fermi surfaces, phonons may couple to singularities in the electronic density of states, which appear at specific wavevectors. These singularities can produce sharp features in phonon dispersions called Kohn anomalies. [3] These typically become stronger with reduced temperature, due to the sharpening of the Fermi surface. A classic example of this behavior is one-dimensional conductors such as $K_2Pt(CN)_4Br_{0.30} \cdot 3D_2O$ (KCP). Due to one-dimentionality of the electronic states, these systems are characterized by Fermi surface nesting at $2\mathbf{k}_f$ ($\mathbf{k}_f$ is Fermi momentum). This nesting greatly enhances the number of possible electronic transitions at $2\mathbf{k}_f$ compared to other wavevectors, which results in softer and broader phonons. For this reason acoustic phonons in KCP show pronounced dips at $\mathbf{q}=2\mathbf{k}_f$ (**Q** is the total wavevecor, **q** is reduced wavevector.)

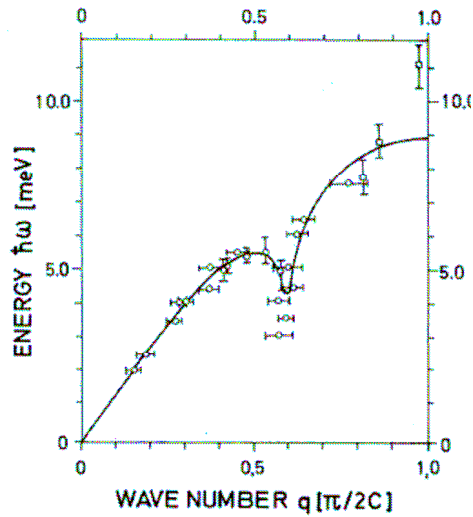

Figure 1. LA phonon branch of $K_2Pt(CN)_4Br_{0.30} \cdot 3D_2O$ in the [001] direction at room temperature. Solid line represents the result of a calculation based on the simple free-electron model as discussed in Ref. 4.

The amount of this phonon softening and broadening depends on the details of the interaction and varies greatly between different systems with Kohn anomalies. Often the broadening is much smaller than the experimental resolution, thus only the softening appears in the experiment.

## 1.3 ab-initio calculations and the role of the q-dependence of electron-phonon matrix element.

It was noticed early on that Fermi surface nesting alone cannot adequately explain phonon softening resembling Kohn anomalies in many cases. For example dispersions of certain phonons in NbC and TaC dip relatively sharply and strongly at wavevectors where the FS nesting is relatively weak. [5,6] These could be modeled by very long-range repulsive interactions or by adding an extra shell with attractive interactions. [6]

S.N. Sinha and B.N. Harmon [7] introduced **q**-dependent electron-phonon coupling to explain these effects. They included **q**-dependent dielectric screening into their model and obtained good agreement with experiment for certain values of adjustable parameters.

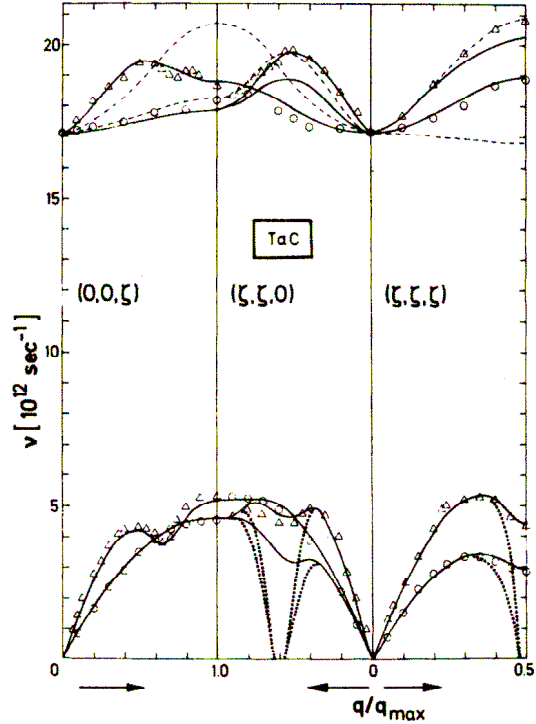

Figure 2. Phonon dispersions of TaC (data points) [5] and calculations based on the double-shell model of W. Weber. [6] Note that the dispersion dip at **Q**=(0,0,0.6) does not correspond to Fermi surface nesting, but originates from **q**-dependence of electron-phonon coupling strength.

Further theoretical development led to ab-initio calculations based on the density functional theory (DFT). These can very accurately predict phonon dispersions and linewidths in many systems without adjustable parameters. [8] However, it is very difficult to isolate contributions of specific electronic states to the phonon self-energy using this approach.

R. Heid et al. made an attempt to overcome this problem in Ru. [9] First they performed both the ab-initio DFT calculations and detailed measurements on a high quality single crystal. The phonon dispersions in Ru had pronounced softening near the M-point, which was well reproduced by theory. The calculated softening became much weaker when electron-phonon coupling was made **q**-independent. This result indicated that both the Fermi surface nesting and the **q**-dependence of electron-phonon matrix element were important (the latter more than the former). The phonon softening disappeared entirely upon exclusion of conduction electrons from the calculation. In addition, the entire dispersion hardened substantially, which may be an artifact of the procedure used to leave out the conduction electrons. This method showed that the phonon dispersion dips in Ru originate from coupling to conduction electrons, but the bare dispersions obtained by excluding them were somewhat arbitrary. This study illustrates the difficulty in separating the bare dispersion from the real part of the phonon self-energy. It is not important for reproducing or predicting phonon anomalies, but is important for making connections with other experiments.

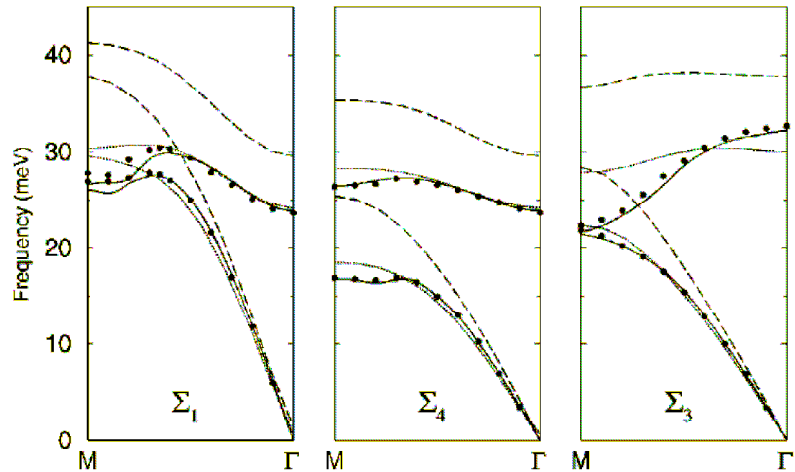

Figure 3. Phonon dispersions in Ru. (from [9]) The data points correspond to experimental results. The solid lines represent results of the full DFT/LDA calculation. Dotted lines were obtained by replacing **q**-dependent electron-phonon matrix element with the average value (which was **q**-independent). The dashed lines represent the DFT calculation without including the conduction electrons. Disappearance of the phonon softening near the M-point upon exclusion of conduction electrons demonstrates that the dip results from electron-phonon coupling.

**1.4 Phonon anomalies in conventional superconductors**

There are two interesting electron-phonon effects observed in measurements of phonon dispersions and linewidths in conventional superconductors. One is the normal state anomalies whose study can elucidate, which phonons contribute the most to the mechanism of superconductivity. The other is the effect of the superconducting gap $2\Delta$ on the phonon spectra below $T_c$, which can be used as a probe of the superconducting gap.

In $MgB_2$ the high superconducting transition temperature, $T_c$, is explained by strong electron-phonon coupling of $E_g$ modes around 80meV near the zone center. A strong dip in the dispersion of these phonons observed in experiments and reproduced by LDA calculations is a clear signature of this coupling. [10,11]

The ab-initio calculations based on DFT/LDA can predict all physical properties of materials that depend on electronic band structure, phonon dispersions and electron-phonon coupling without adjustable parameters. In particular, they can be used to calculate $T_c$ based on Migdal-Eliashberg theory. [12] Several ab-initio calculations of phonon dispersions as well as $T_c$s were performed for the transition metal carbides and nitrides in order to explain relatively high transition temperatures in some and not in others. [13,14,15] They correctly reproduced phonon dispersions including the anomalies discussed in Sec. 1.3, [13,15,16] and established a correlation between the phonon anomalies and $T_c$. They also suggested that the phonon anomalies are associated with the Fermi surface nesting [15], but more detailed calculations along the lines of Ref. [9] need to be performed to separate the role of nesting from the enhancement of the electron-phonon matrix element. One interesting possibility that may need to be explored, is that additional screening near the nesting wavector may enhance the electron-phonon matrix elements, thus the Kohn anomalies at the nesting wavevectors may be stronger than expected from enhanced electronic response due to nesting alone.

In a conventional superconductor with $T_c$=15K, $YNi_2B_2C$, the transverse acoustic phonons near **q**=(0.5 0 0) soften and broaden on cooling [17] as expected from electron-phonon coupling. W. Reichardt et al. calculated phonon frequencies and linewidths using LDA. [18] These calculations reproduced this phonon anomaly and correctly predicted an additional wavevector (**q**=0.5 0.5 0) where acoustic phonons couple strongly to conduction electrons. [19] The energies of both soft phonons are comparable to the superconducting gap energy. In this case the opening of the superconducting gap has a strong influence on the phonon spectral function.

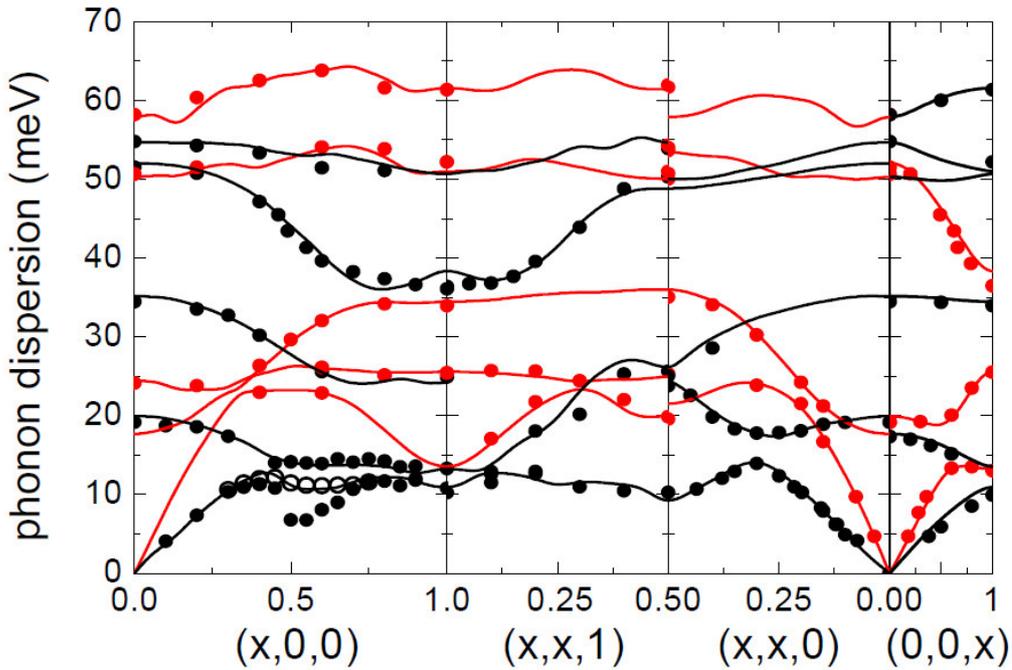

Figure 4. Calculated (lines) and observed (filled dots) phonon frequencies in $YNi_2B_2C$ at 20K. Open dots along [100] were measured at 300 K. Branches shown in red/black refer to phonons of predominantly longitudinal/transverse polarization respectively. The horizontal axes denote different crystallographic directions in reciprocal lattice units (r.l.u.). The theoretical results were scaled up by a factor of 1.03 (from Ref. [19]).

These phonons were so broad, that their normal state spectra extended below the low temperature superconducting gap 2Δ. In this case, when the 2Δ opens in the electronic spectrum in the superconducting state, the damped harmonic oscillator approximation of the phonon breaks down. P.B. Allen et al. [20] developed a theory precisely for this case, which predicted that phonon lineshapes in the superconducting state should contain either a step or a sharp feature very close to 2Δ depending on the values of the phonon energy, electron-phonon coupling and 2Δ. Normal state lineshape fixes the first two parameters, which leaves only one adjustable parameter, i.e. the superconducting gap. Detailed measurements of F. Weber et al. [21] confirmed this theory. (Fig. 5) They also suggested how to use phonon measurements to determine the magnitude of 2Δ and to probe gap anisotropy.

The effects of superconductivity on phonons in the cuprates were discussed in detail in Ref. [1] and will not be reviewed here.

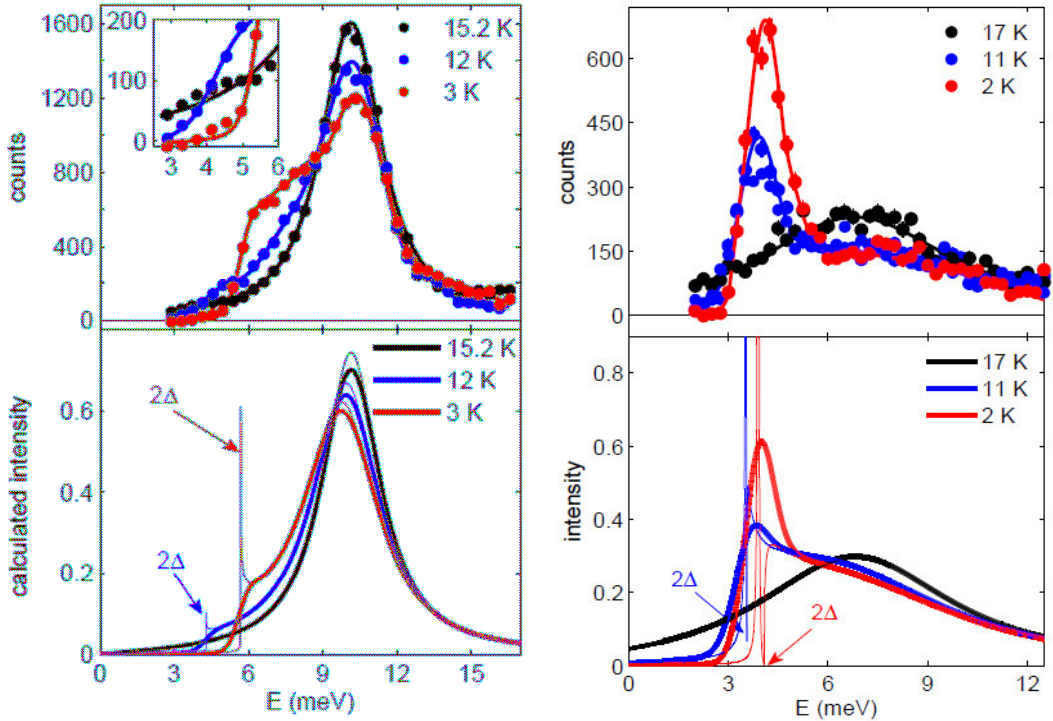

Figure 5. Top panels: Evolution of the neutron-scattering profile measured on YNi$_2$B$_2$C at **Q** =(0.5,0.5,7) (left) and **Q** =(0.5,0,8) (right) above and below T$_c$=15K. Bottom panels: Thin lines represent calculated phonon lineshapes based on the theory of Allen et al. [20], using parameters extracted from the line-shape observed in the normal state. The thick lines are obtained after convolution with the experimental resolution. (from Ref. [21])

## 1.5 Phonon anomalies with and without the Fermi surface nesting in Chromium

A different situation appears in Chromium where the Fermi surface nesting is responsible for an incommensurate spin density wave (SDW) at a nesting wavevector **q**$_{sdw}$=(0.94,0,0) and, as a secondary effect, of the charge density wave (CDW) at **q**$_{cdw}$=(0.11,0,0). [22] W. M. Shaw and L. D. Muhlestein measured phonon dispersions in chromium by INS. [23] They reported soft phonons around **q**=(0.9,0,0), which is near **q**$_{sdw}$, as well as near **q**=(0.45,0.45,0) where some nesting has also been calculated, although the data were not accurate enough to establish their exact wavevectors.

Recently Lamago et al. [24] performed a more precise and comprehensive set of measurements by IXS, which showed that the two anomalies actually appear at **q**$_{sdw}$ and **q**=(0.5,0.5,0) respectively. A surprising result was that a transverse phonon branch softened throughout the zone boundary between **q**=(0.5, 0.5, 0) and (1, 0, 0), i.e. for **q**=(0.5+h,0.5-h,0), where 0<h<0.5. LDA-based calculations performed as a part of this investigation, successfully reproduced the observed phonon softening. (Fig. 6) However, the electronic response function that couples to the phonons obtained from the same calculations showed no clear features corresponding to the phonon dips along the zone boundary. Thus these phonon dips come exclusively from the **q**-dependence of the electron-phonon matrix element. To the best of my knowledge, this is the first clear observation of a phonon softening in a narrow **q**-range exclusively due to the **q**-dependence of the electron-phonon matrix element.

This result was further corroborated by the effect of numerical smearing of the Fermi surface on the calculated phonon dispersions. The smearing suppressed only the calculated effect at **q**=(0.94,0,0), but not at **q**=(0.5+h,0.5-h,0) for any h including h=0. (Fig. 6 a,b) Such a smearing makes it possible to differentiate between the effects of Fermi surface nesting and electron-phonon coupling enhancement because it reduces the former but not the latter. D. Lamago et al. concluded that only the softening at **q**$_{sdw}$ originates from Fermi surface nesting. The effect at **q**=(0.5+h,0.5-h,0) comes exclusively from the **q**-dependence of electron-phonon coupling. This is true even for h=0, which is near a nesting feature previously thought [23] to be responsible for phonon softening. In fact it makes a negligible contribution to the phonon self-energy, since the nesting feature disappears even at small h, but the phonon renormalization does not become smaller in either the calculation or the experiment. [24].

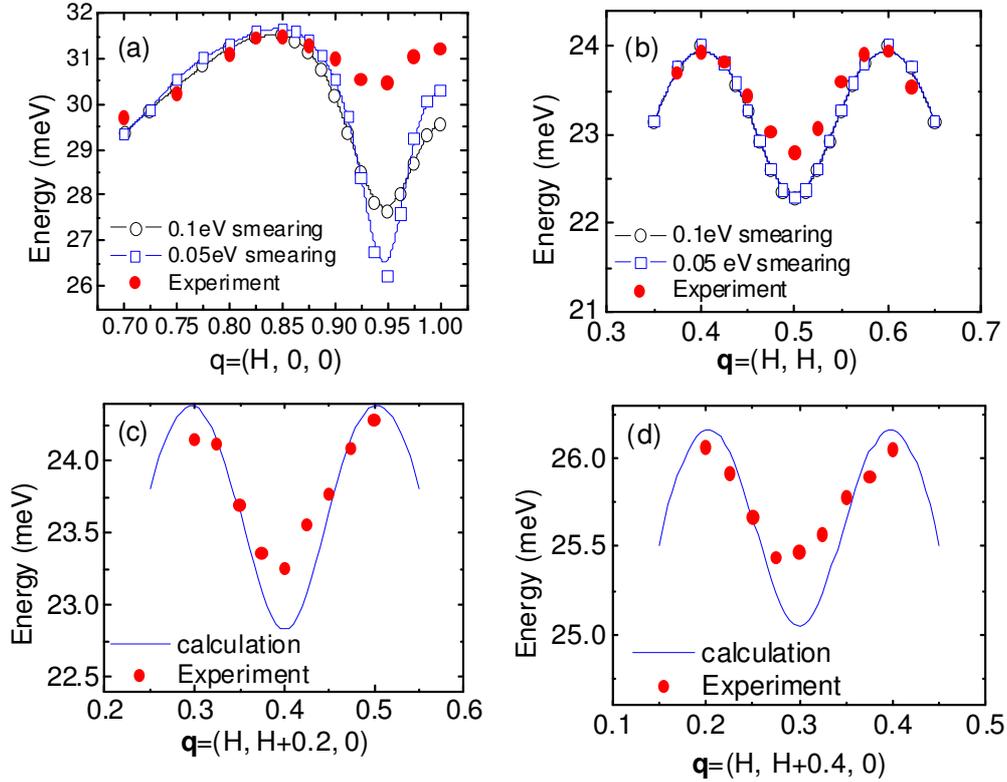

Figure 6. A comparison between measured (red) and calculated (blue, black) phonon dispersions and the LDA calculation in Cr. Only the data for the lowest energy transverse phonons near the zone boundary line connecting **q**=(1,0,0) and **q**=(0.5,0.5,0) are shown. The smearing has a strong effect in (a) but not in (b). (from Ref. [24])

**1.6 Lessons learned from conventional metals**

It follows from the investigations reviewed above as well as from other similar work that conventional metals including phonon-mediated superconductors can be understood in terms of DFT in the approximations that ignore electron-electron interactions, except when mediated by phonons. These approximations work remarkably well in a variety of metals. They correctly reproduce electron-phonon effects, and account qualitatively for the variations in superconducting $T_c$s in compounds of similar structure and chemistry. The rest of this article will show that these approximations fail completely for phonons in the cuprates, most likely, because electronic correlations play a key role in brining about experimentally observed strong electron-phonon coupling in the high $T_c$ superconductors.

## 2. Phonon anomalies in doped $La_2CuO_4$
## 2.1 Summary of early work

Calculations performed soon after the discovery of high temperature superconductors suggested that electron-phonon coupling is too weak to account for high temperature superconductivity. [25,26] However, measurements performed in conjunction with shell model calculations showed that the bond-stretching branch softened strongly towards the zone boundary as doping increased from the insulating phase to the superconducting phase. [27] This behavior pointed towards strong electron-phonon coupling and a possible role of the zone boundary "half breathing" bond-stretching mode in the mechanism of high temperature superconductivity. [28] It later became apparent that this trend continues into the overdoped nonsuperconducting phase, which indicates that zone boundary softening is related to the increase of metallicity with doping rather than to the mechanism of superconductivity (Fig. 7). [29]

These experiments and related calculations are extensively covered in the previous review by L. Pintschovius. [1] Here I will focus not on the zone boundary, but half-way to the zone boundary in the [100]- direction (along the Cu-O bond) where the most interesting physics has been observed. This work began with the INS experiments of McQueeney et al. [30] who reported anomalous lineshape and temperature dependence of the bond-stretching phonons in $La_{1.85}Sr_{0.15}CuO_4$ near **q**=(0.25, 0, 0) and interpreted these results in terms of line splitting due to unit cell doubling. (Fig. 8)

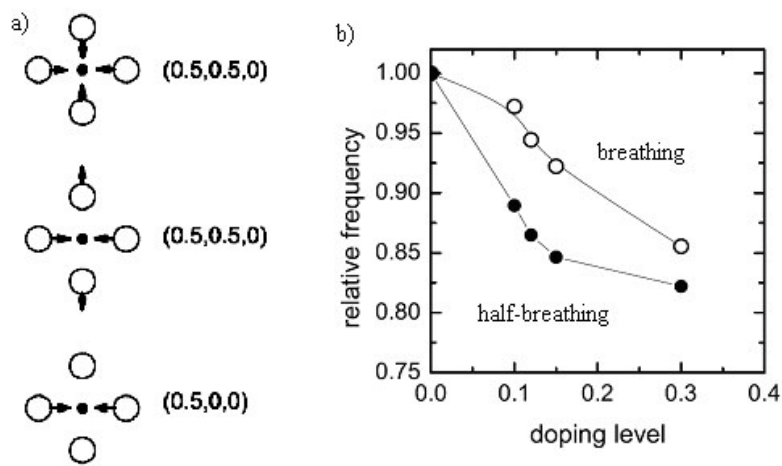

Figure 7. a) Displacement patterns of zone-boundary bond-stretching modes in cuprates. Top: longitudinal mode in the [110]- direction (breathing mode); middle: transverse mode in the [110]-direction (quadrupolar mode); bottom: longitudinal mode in the [100]-direction (half-breathing mode). Circles and full points represent oxygen atoms and copper atoms, respectively. Only the displacements in the Cu-O planes are shown. All other displacements

are small for these modes. b) Schematic of doping dependence of the breathing **q**=(0.5,0.5,0) and half-breathing **q**=(0.5,0,0) zone boundary mode frequencies. This behavior is probably not directly related to the mechanism of superconductivity, since the softening continues into the overdoped nonsuperconducting part of the phase diagram. From Ref. [29]

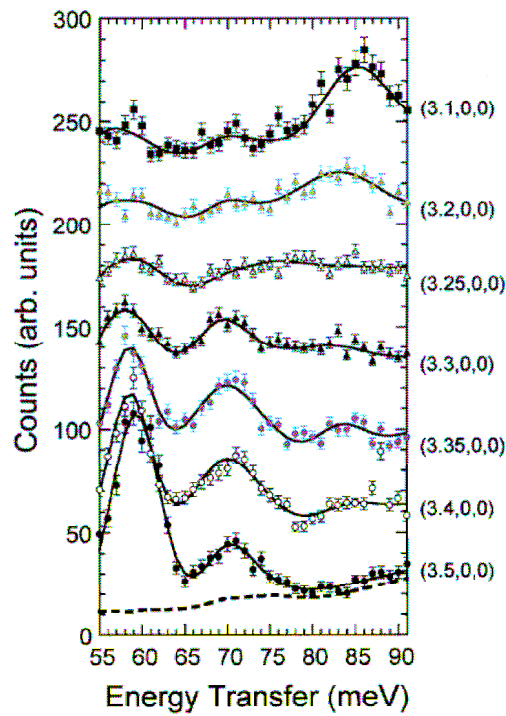

Figure 8. Bond-bending (around 60meV) and bond-stretching (above 65meV) branches in optimally-doped LSCO. [30]

This interpretation evolved considerably in recent years as a result of further measurements and calculations. Pintschovius and Braden repeated the experiment using different experimental conditions, which had a higher energy resolution due to their use of Cu220 monochromator. [31] They also measured the interesting wavevector range between q=0.1 and 0.4 in the so-called focusing condition with the tilt of the resolution ellipsoid matching the phonon dispersion, which further improved the resolution compared to Ref. [30] (see sec. 2.2 for a more detailed discussion) Pintschovius and Braden reported enhanced linewidth near the same wavevector (with the strongest broadening at **q**=(0.3, 0, 0)), but did not see any splitting of the phonon line.

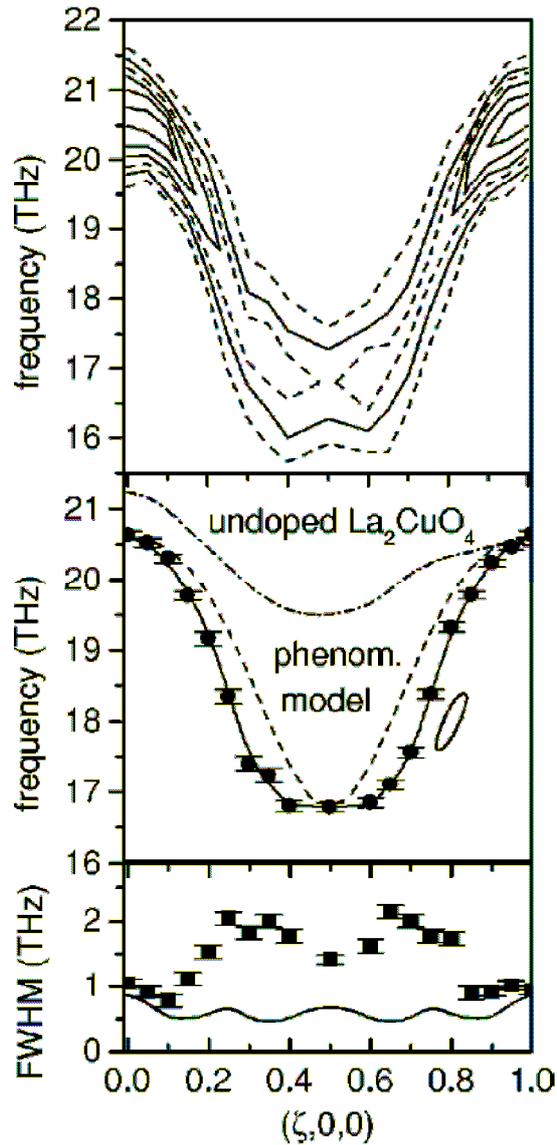

Figure 9. Results of measurements on $La_{1.85}Sr_{0.15}CuO_4$ performed with a better energy resolution and similar in-plane wavevector resolution than in Fig. 8. [31]

The origin of these effects was not clear at the time, but phonon anomalies near half way to the zone boundary suggested a possible connection to incipient stripe formation, i.e. nanoscale phase separation characterized by charge-rich lines separating charge poor antiferromagnetic domains. [32,33,34]

D. Reznik et al. investigated the same bond stretching branch in $La_{1.875}Ba_{0.125}CuO_4$ where static stripes appear at low temperatures. [35] They performed the first set of measurements in the same scattering geometry as Pintschovius and Braden (at wavevectors (5-4.5,0,0) or (5-4.5,0,1)) and found that the phonon dispersion could be described with two components: One had a "normal" dispersion following a cosine function (blue line), and the other softened and broadened abruptly at $q=(0.25, 0, 0)$ (black line). The possible relationship between the phonon anomaly and stripe formation is further explored in Sec. 3.5.

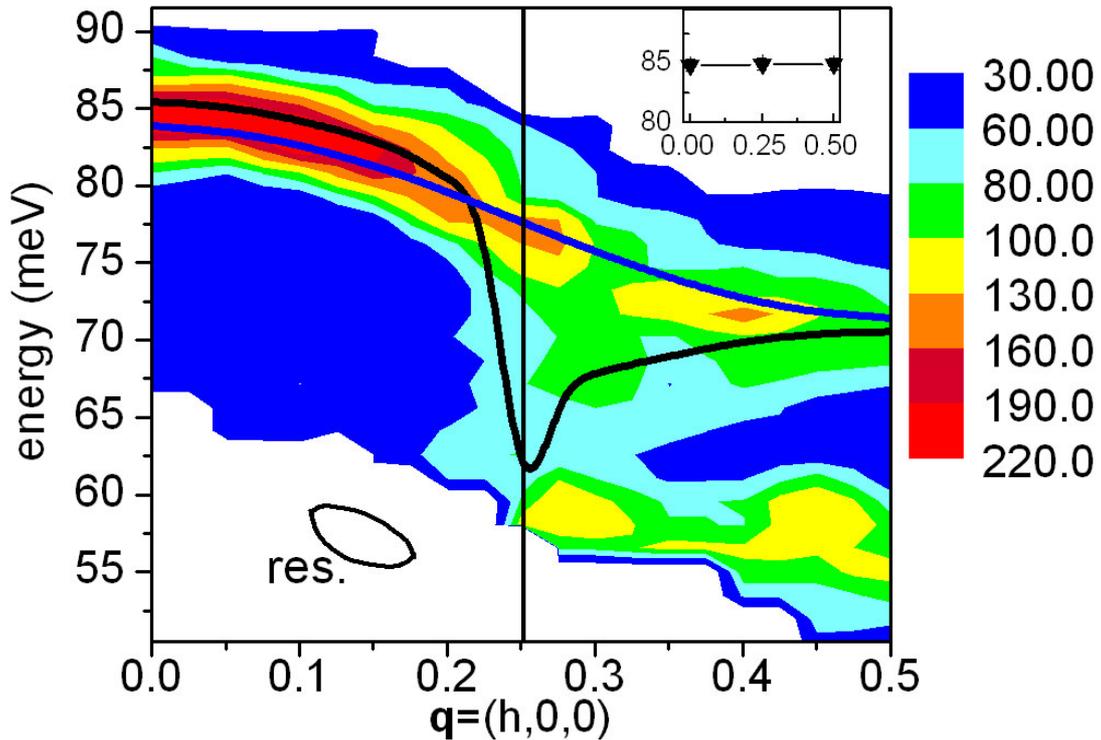

Figure 10. (a) Color-coded contour plot of the phonon spectra observed on $La_{1.875}Ba_{0.125}CuO_4$ at 10 K. The intensities above and below 60 meV are associated with plane-polarized Cu-O bond-stretching vibrations and bond-bending vibrations, respectively. Lines are dispersion curves based on two-peak fits to the data. The white area at the lower left corner of the diagram was not accessible in this experiment. The ellipse illustrates the instrumental resolution. The inset shows the dispersion in the [110]-direction. The vertical line represents the charge stripe ordering wave vector. Blue/black line represents the "normal"/"anomalous" component respectively in the original interpretation of the authors. Subsequent work showed that a substantial part of the intensity in the "normal" component is an artifact of finite wavevector resolution in the transverse-direction (out of the page).

D. Reznik et al. [35,36] found that there was an overall hardening of the spectral weight on heating in both $La_{1.875}Ba_{0.125}CuO_4$ and $La_{1.85}Sr_{0.15}CuO_4$ at **q**=(0.25,0,0). (see Fig. 11) This indicates that the anomalous broadening does not originate from anharmonicity or structural inhomogeneity, since these have the opposite or no temperature dependence.

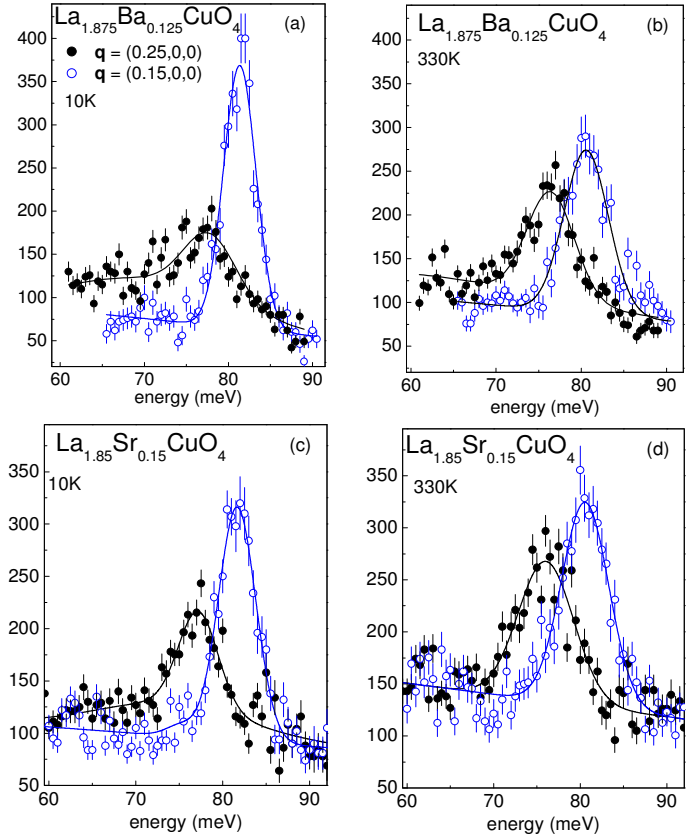

Figure 11. Temperature dependence of the bond-stretching phonons at select wavevectors. Energy scans taken on $La_{1.875}Ba_{0.125}CuO_4$ (a,b) and on $La_{1.85}Sr_{0.15}CuO_4$ (c,d,e) at 10K (a,c) and 330K (b,d) (Ref. 36). The phonon at **q**=(0.15,0,0) is "normal" in that it has a Gaussian lineshape on top of a linear background. This background results from multiphonon and incoherent scattering and has no strong dependence on **Q**. The intensity reduction of this phonon in $La_{1.875}Ba_{0.125}CuO_4$ from 10K (a) to 330K (b) is consistent with the Debye-Waller factor. At **q**=(0.25,0,0), there is extra intensity on top of the background in the tail of the main peak. It originates from one-phonon scattering that extends to the lowest investigated energies, while the peak intensity is greatly suppressed as discussed in the text. The effect is reduced but does not disappear at 330K. Note that 330K in (b) is shown instead of 300K in the same plot in Ref. 36 because of a typographical error in the latter. Integrated intensity of the phonon decreases from **q**=(0.15,0,0) to **q**=(0.25,0,0) due to the decrease of the structure factor.

Pintschovius and Braden [31] observed no temperature dependence at q=(0.3,0,0) (it is equivalent to **q**=(0.7, 0, 0) if interlayer interactions are neglected) although the zone center phonon of the same branch softened on heating. This softening is due to increased anharmonicity and should affect the entire branch. Absence of softening at **q**=(0.3,0,0) that they report, indicates that there is a counterbalancing trend, which makes their results agree qualitatively with Refs. [35,36]. I will explain the reason for the quantitative difference in the following section.

McQueeney et al [30] reported suppression of the anomalous behavior at room temperature. However, the room temperature data of [30] suffered from a much stronger background than the low temperature data and relatively large statistical error. Ref. 36, which had a much better resolution and signal-to-background ratio, but was limited to only three wavevectors, also reported a suppression of the anomalous behavior at 330K. In this regard the two studies are consistent, although Ref. 30 claims a much more radical change of the phonon dispersion than reported in Ref. 36. To resolve this disagreement it is necessary to perform measurements covering the entire BZ at 300K with the experimental configuration of Ref. 36.

## 2.2 Recent IXS Results.

Neutron scattering experiments have a relatively poor **Q** resolution. For the cuprates its full width half maximum (FWHM) is on the order of 15% of the in-plane Brillouin zone. The effects of finite **Q** resolution in the longitudinal direction have been carefully considered in early studies, but the finite resolution in the transverse direction has not. In this section I will discuss recent IXS work and will show that some previous experiments need to be reinterpreted taking into account the finite transverse **Q**.

More recent measurements using both INS [36] and IXS [37] yielded a somewhat surprising result that the anomalous softening/broadening for **q**=(0.25,k,0) occurs only very close to k=0. For example in $La_{1.84}Nd_{0.04}Sr_{0.12}CuO_4$ the phonon anomaly significantly weakened at |k|≈0.08 compared to k=0, and disappeared entirely at |k|=0.16 (see Fig.12). [37] The "normal" component for k≈0 was significantly suppressed.

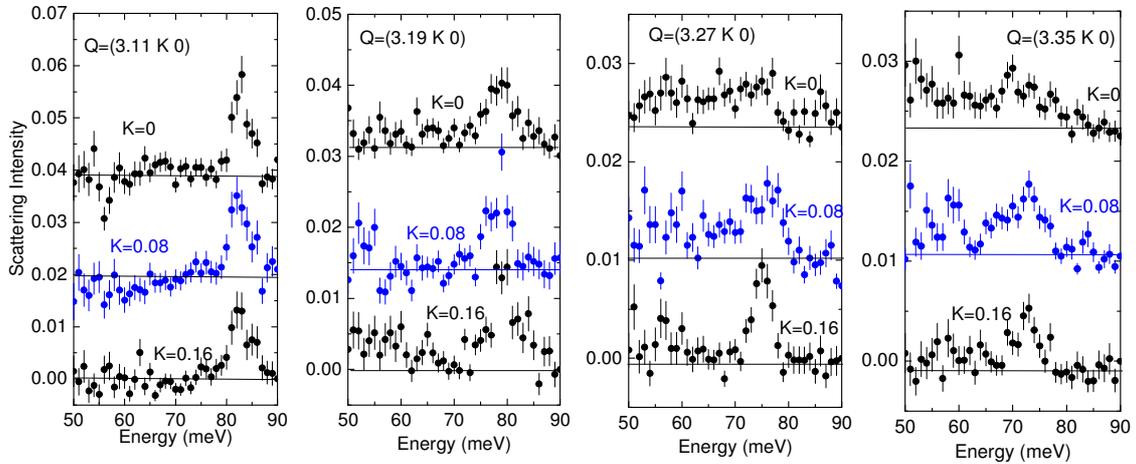

Figure 12. IXS energy scans after subtraction of the elastic tail and a constant term corresponding to the stray radiation. The scans were taken with **q**=(h,k,0) with h = 0.11, 0.19, 0.27, 0.35 (from the left to the right column) and k as indicated in the figure. The most interesting features are the suppression of the two-component behavior seen by INS at 77 meV near **Q**=(3.27,0,0) and the rapid narrowing and hardening of the phonon line from k=0 to k=0.16 for **Q**=(3.27,k,0). From Ref. [37]

Neutron measurements have a much lower resolution in the k-direction, i.e. INS experiments nominally performed with k=0 include a significant contribution from wavevectors with |k|>0.08 even in the most optimal configuration (a-b scattering

plane). Thus most of the intensity in the "normal" component in the INS measurements probably comes from these phonons with |k|>0.08.

With this information it now becomes possible to explain why the temperature effect in [31] was weaker than in [36]. The experiment of Pintschovius and Braden [31] was performed in the a-c scattering plane, which had poorer wavevector resolution in the k-direction, whereas the other study was performed in the a-b scattering plane, which had a better k-resolution. [38] Since the phonon anomaly is sharp in the k-direction, the anomalous behavior should be masked by the "normal" phonons with |k|>0 in the a-c scattering plane more than in the a-b scattering plane. This "masking" would also reduce the phonon linewidth in [31] compared to the measurement in [36] performed with better k-resolution.

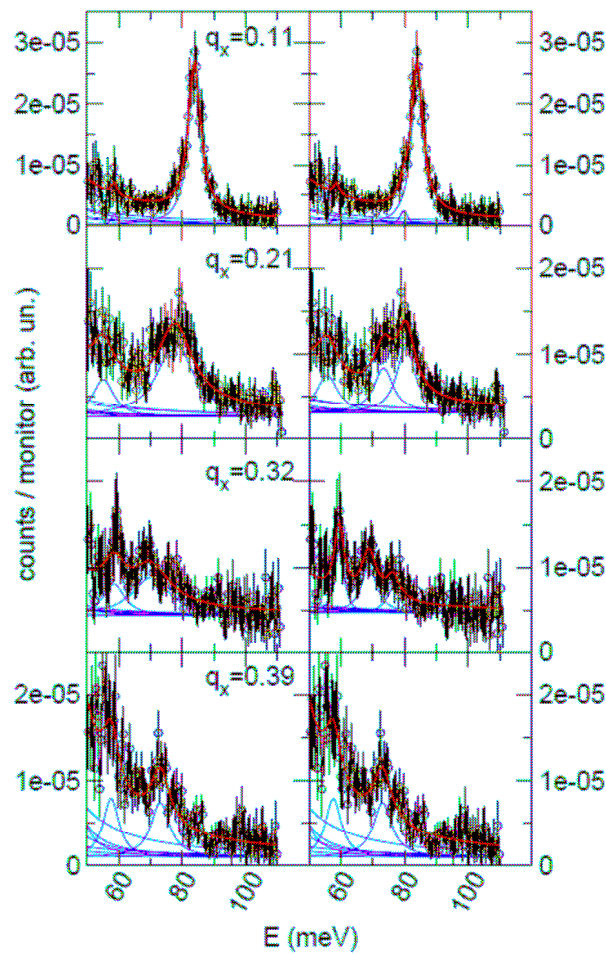

Figure 13: Inelastic X-ray scattering spectra of $La_{1.86}Ba_{0.14}CuO_{4+\delta}$, at $\mathbf{Q}=(3+q_x, q_y, 0)$ ($q_y<0.04$). In the left column a single peak is used to fit the data. In the right column the data are fitted with two Cu-O bond stretching modes. (from Ref. [39])

D'Astuto et al. [39] reported two-branch behavior in IXS spectra of $La_{1.86}Ba_{0.14}CuO_4$ with clearly resolved "normal" and "anomalous" components. (Fig. 13) This result seems to be different from the $\mathbf{Q}=(3.27, 0, 0)$ data of Fig. 12 measured also by IXS with a much higher energy resolution, as well as with the results of J. Graf et al. on $La_{1.92}Sr_{0.08}CuO_4$ [40], which are consistent with either a single broad peak or two strongly overlapping peaks. It is possible that since D'Astuto et al. measured a Ba-doped sample, and D. Reznik et al. investigated the Sr-doped systems, the difference may come from Ba vs. Sr doping. It is necessary to perform further experiments to clarify this potentially important issue.

**2.3 Doping dependence**

Fig. 14 shows the bond-stretching phonon dispersion and linewidth for $La_{2-x}Sr_xCuO_4$ with x=0.07,0.15, and 0.3. The dispersion is compared with the cosine function that typically comes out of DFT calculations (see for example [41,42,43]). Here the data presented in Fig. 4 of Ref. 35 are combined with some unpublished results and refitted using the model that includes all phonon branches picked up by the spectrometer resolution as opposed to gaussian peaks. [44] Such an analysis provides more accurate values of intrinsic phonon linewidths. The strongest dip below the cosine function and the biggest peak of the linewidth is observed at optimal doping where the $T_c$ is highest. These are smaller at x=0.07 and disappear in the overdoped nonsuperconducting sample with x=0.3. The position of the phonon anomaly does not change with doping.

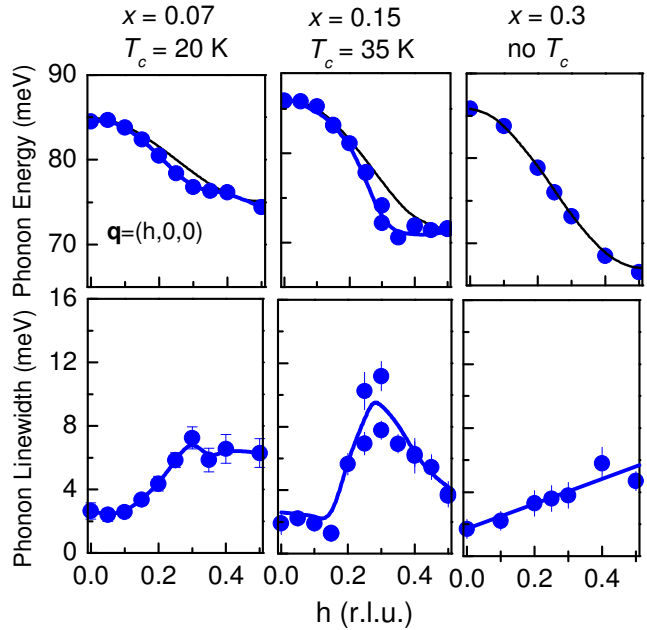

Figure 14. Bond-stretching phonon dispersion (top row) and linewidth (bottom row) at three doping levels. Black lines represent downward cosine dispersion. The overall increase in the linewidth of the bond-stretching mode towards the zone boundary appears to be doping-independent. Softening compared with the cosine dispersion as well as the linewidth enhancement half-way to the zone boundary do not shift with h between x=0.07 and 0.15.

Comparison with the overdoped sample, where the physics are conventional, allows to identify the effects of electron-phonon coupling that are intrinsic to optimal doping. Figure 15 shows the schematic of the anomalous phonon broadening that appears on top of the broadening observed in the x=0.3 sample. The anomalous broadening peaks at $\mathbf{q}=(0.3,0,0)$ and weakens rapidly in the longitudinal and transverse directions.

This effect is phenomenologically very similar to the renormalization of the acoustic phonons at specific wavevectors discussed in section 1. Next, I will show that profound differences exist between La$_{2-x}$Sr$_x$CuO$_4$ and conventional metals.

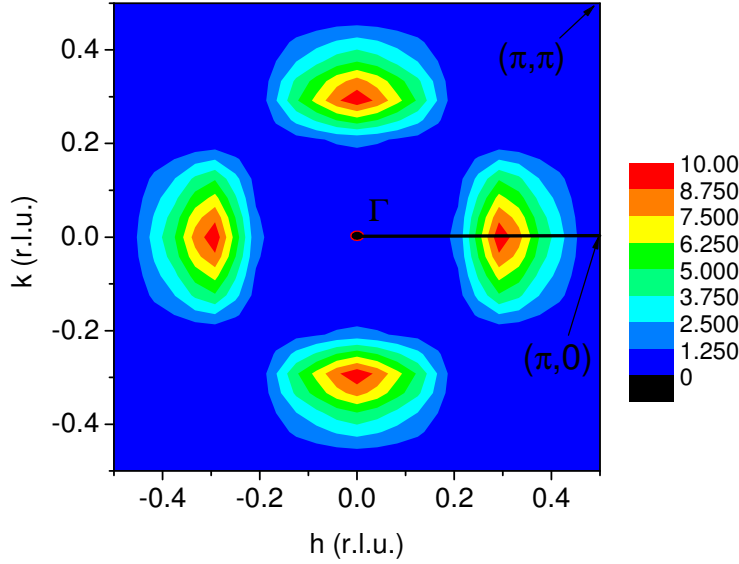

Figure 15. Qualitative picture of the difference in the linewidths of the bond-stretching phonon in optimally-doped (x=0.15) and overdoped (x=0.3) La$_{2-x}$Sr$_x$CuO$_4$ as a function of wavevector in the ab-basal plane based on [29] and [36]. The units are meV. The solid line indicates the [100] direction along which most of measurements were performed.

**2.4 Comparison with density functional theory**

As discussed in Sec. 2, density functional theory gives a good description of phonon dispersions in metals where electron-electron interactions can be neglected. Giustino et al. [42] performed such a calculation in the generalized gradient approximation (GGA) for La$_{1.85}$Sr$_{0.15}$CuO$_4$, which approximately agreed with the early experimental data of [30]. In particular, they reproduced the overall downward dispersion of the longitudinal bond-stretching branch. However, the strong effect in the bond-stretching phonon half way to the zone boundary was not apparent in figure 1 of Ref. [42], because the 300K results and 10K results were plotted together. In the brief communiation arising from the article of Giustino et al, D. Reznik et al. showed that DFT did not reproduce the phonon anomaly half way to the zone boundary that appears in Ref. [30,31,35,36,37,39] as well as in later experiments discussed above in detail. (Fig. 16, [45]).

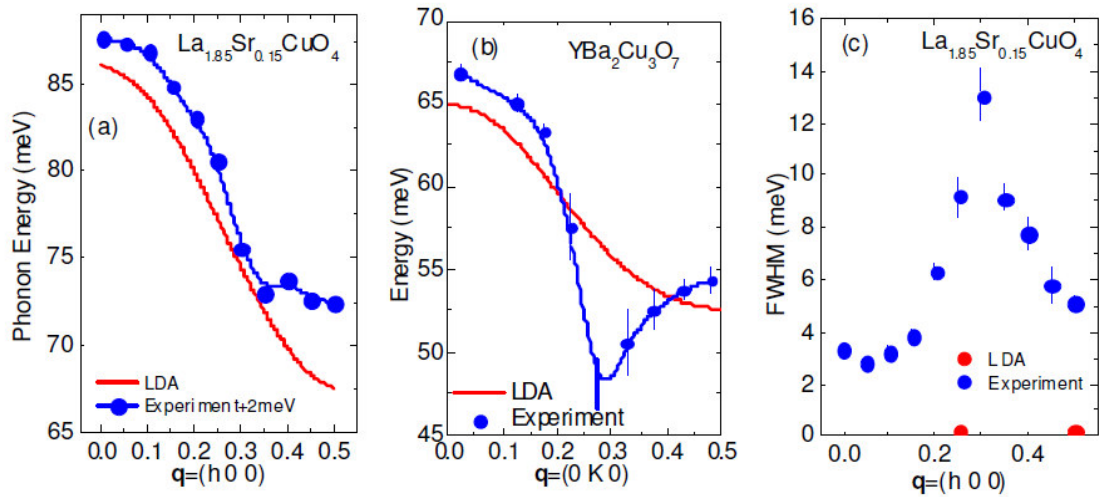

Figure 16. Comparison of some LDA predictions with experimental results for $La_{1.85}Sr_{0.15}CuO_4$ [42] and $YBa_2Cu_3O_7$ [74] at 10K. (a,b) Experimental bond-stretching phonon dispersions compared to LDA results. The data in (a) are shifted by 2meV. (c) Phonon linewidths in $La_{1.85}Sr_{0.15}CuO_4$ compared with LDA results on $YBa_2Cu_3O_7$. Ref. [42] contains no linewidth results for $La_{1.85}Sr_{0.15}CuO_4$ but they should be similar.

Giustino et al. also calculated the phonon contribution to the kink in the electronic dispersion observed by photoemission spectroscopy. (Fig. 17) Such kinks may result from bosonic modes interacting with electrons. One of the proposed mechanisms of high temperature superconductivity is Cooper pairing via this boson, thus it is important to identify its origin. Fig. 17 shows that in DFT phonon contribution to the kink is five times smaller than observed. Analogous results for both the phonon dispersions [43] and the kink [46] were obtained earlier for $YBa_2Cu_3O_7$.

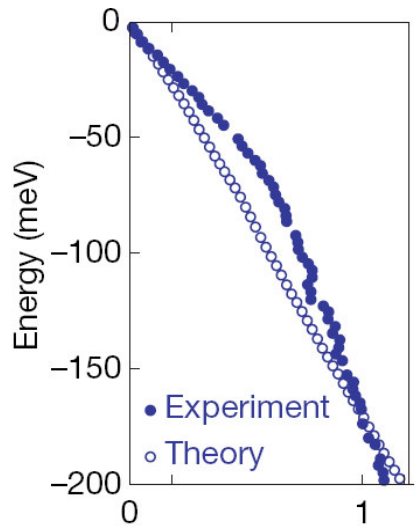

Figure 17. Comparison between the quasi particle dispersion relations obtained from the peaks of the photoemission spectra and calculations of Ref. [42], including the renormalization due to the electron–phonon interaction in $La_{1.85}Sr_{0.15}CuO_4$.

However, D. Reznik et al. argued that the failure of the DFT to reproduce the phonon anomaly indicates that it is possible that electronic correlations ignored by the DFT enhance electron-phonon coupling. This enhancement could result in a much stronger kink than calculated by the DFT. It is interesting that many-body calculations predict a substantial enhancement of the coupling to bond-stretching phonons compared to DFT. (see for example refs. 28 and 47) t-J model-based calculations describe interesting doping dependence of the zone boundary phonons, suggesting that strong correlations might be relevant. [48,49] Recent high resolution photoemission measurements have found an isotope effect in the dispersion kink, hinting at an important role of phonons. [50]

This connection is reinforced by the fact that deviations of both the electronic dispersion and of the phonon dispersion from DFT predictions disappear in overdoped nonsuperconducting $La_{2-x}Sr_xCuO_4$ as shown in Fig. 18.

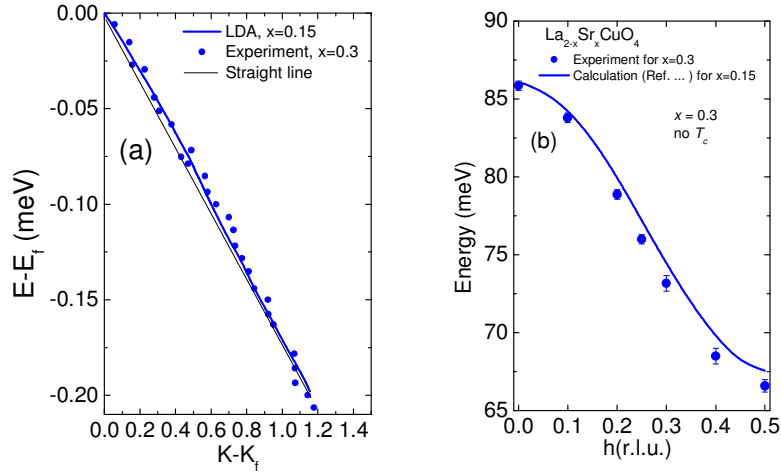

Figure 18. Comparison of the photoemission kink (a) and phonon dispersion (b) in $La_{1.7}Sr_{0.3}CuO_4$ with the DFT calculations for $La_{1.85}Sr_{0.15}CuO_4$. (based on results of [29,42,51]). Black thin line in (a) is a straight line serving as a guide to the eye.

## 2.5 Connection with stripes and other charge-inhomogeneous models

The bond-stretching phonon anomaly is strongest in $La_{1.875}Ba_{0.125}CuO_4$ and $La_{1.48}Nd_{0.4}Sr_{0.12}CuO_4$, compounds that exhibit spatially modulated charge and magnetic order, often called stripe order. It appears when holes doped into copper-oxygen planes segregate into lines, which act as domain walls for an antiferromagnetically ordered background. Static long-range stripe order has been observed only in a few special compounds such as $La_{1.48}Nd_{0.4}Sr_{0.12}CuO_4$ and $La_{1.875}Ba_{0.125}CuO_4$ where anisotropy due to the transition to the low temperature tetragonal structure provides the pinning for the stripes while superconductivity is greatly suppressed. [34] In contrast, the more common low temperature orthorhombic (LTO) phase does not provide such a pinning and static stripes do not form. In the LTO phase the stripes are assumed to be purely dynamic, which makes their detection extremely difficult. [52] Here I discuss the possible relation between the phonon anomaly and dynamic stripes.

A detailed comparison between the bond-stretching phonon dispersion in stripe-ordered compounds and optimally-doped superconducting $La_{1.85}Sr_{0.15}CuO_4$ was performed by D. Reznik et al. [36] They found the strongest phonon renormalization at h=0.25 in the presence of static stripes and h=0.3 at optimal doping. (Fig. 19) It appears that static stripes pin the phonon anomaly at the stripe ordering wavevector.

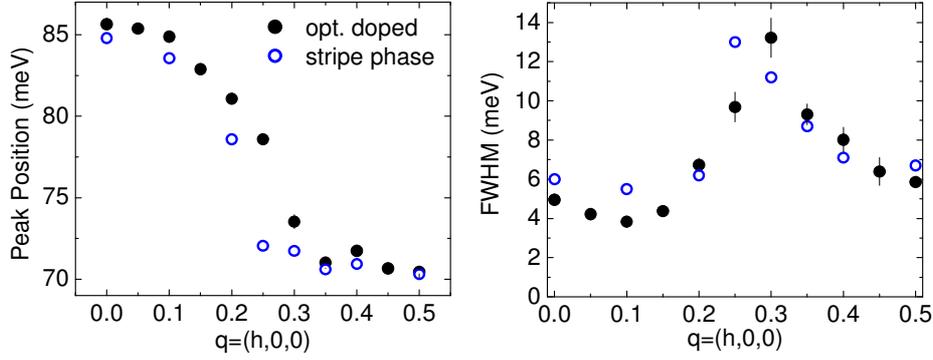

Figure 19. Comparison of the phonon dispersions (a) and linewidth (b) of the bond-stretching branch in $La_{1.85}Sr_{0.15}CuO_4$ and $La_{1.48}Nd_{0.4}Sr_{0.12}CuO_4$. (from Ref. [36])

Two mechanisms of the impact of dynamic stripes on phonons have been proposed: One is that the phonon eigenvector resonates with the charge component of the stripes; The other is that one-dimensional nature of charge stripes makes them prone to a Kohn anomaly, which renormalizes the phonons. In the first scenario (2D picture) the propagation vector of the anomalous phonon must be parallel to the charge ordering wavevector, whereas in the second scenario (1D picture) it must be perpendicular to the charge ordering wavevector. (Fig. 20)

An important clue is that the phonon anomaly disappears quickly as one moves away from k=0 along the line in reciprocal space: **q**=(0.25,k,0) as shown in Refs. [36] and [37] and discussed in section 2.2. Such behavior is expected from the matching of the phonon wavevector and the stripe propagation vector. In contrast, a simple picture of a Kohn anomaly due to 1-D physics inside the stripes predicts a phonon anomaly that only weakly depends on k. This observation favors the 2D picture, but an important caveat is that it may be possible to reconcile the 1D picture with experiment by including a decrease of the electron-phonon matrix element away from k=0. [53]

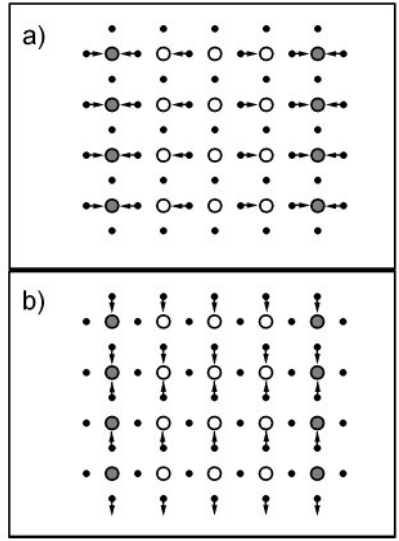

Figure 20. Schematic of the eigenvectors for the phonons with **q**=(0.25 0 0) propagating perpendicular (a) and parallel (b) to the stripes. Open circles represent hole-poor antiferromagnetic regions, while the filled circles represent the hole-rich lines. (from Ref. [36])

Another way to distinguish between the two scenarios is to consider the doping dependence of the wavevector of maximum phonon renormalization, $q_{max}$. In the stripe picture, $q_{max} = 2q_{in}$, ($q_{in}$ is the wavevector of incommensurability of low energy spin fluctuations). [54] At doping levels of x=0.12 and higher, $q_{in}$= 0.125, which gives the charge ordering wavevector of 0.25. This value is indeed close to $q_{max}$. At x=0.07 $q_{in}$= 0.07. This gives the charge stripe ordering wavevector of 0.14 whereas $q_{max}$=0.3. This discrepancy appears to contradict the 2D picture. But again there is a caveat: Anomalous phonons occur at a fairly high energy of about 75 meV, and a comparison to the dynamic stripe wavevector at low energies may not be appropriate.

Thus the question of which picture, 1D or 2D, agrees better with the data is not yet settled.

If the phonon renormalization is driven by static stripes, one may expect to see different behavior for phonons propagating parallel or perpendicular to the stripe propagation vector. [55] In this case the phonon should split into two peaks. The dynamic stripes, according to M. Vojta et al. [56], may not break tetragonal symmetry, because fluctuations can occur in both directions simultaneously. Thus a single-peak anomalous phonon lineshape is compatible with dynamic stripes.

**2.6 Phonon anomalies and superconductivity**

Origin of superconductivity in the cuprates is still hotly debated. A conventional phonon-mediated mechanism dropped out fairly early in this debate in large part because LDA calculations showed that electron-phonon coupling is too weak to explain superconductivity. However, recent measurements have shown (see sec. 2.4) that electron phonon coupling of the bond-stretching mode is much stronger than calculated by LDA, which invalidates this claim. In fact the magnitude of the phonon renormalization in $La_{1.85}Sr_{0.15}CuO_4$ is similar to that of the $E_g$ phonons in $MgB_2$, [57] which is held responsible for superconductivity with an even higher $T_c$. It is also interesting that these $E_g$ phonons have roughly the same energy as the bond-stretching phonons in the cuprates. In addition, the magnitude of the bond-stretching phonon renormalization in $La_{1.85}Sr_{0.15}CuO_4$ is similar to that recently observed in a bilayer manganite where exceptionally strong electron-phonon coupling induces a structural phase transition accompanied by the colossal magnetoresistance. [58]

Can these results be interpreted in favor of a conventional mechanism of superconductivity at least in the $La_{2-x}Sr_xCuO_4$ family where $T_c$ is relatively low?

If the mechanism of superconductivity were conventional, DFT calculations based on Eliashberg formalism with noninteracting electrons (except for the exchange of phonons) should explain the experimental results at least qualitatively. Since this is not the case, in a strict sense the answer is "NO". However, the appearance of the maximum $T_c$ and the strongest bond-stretching phonon anomaly around the same doping points towards some other connection with the mechanism of superconductivity. [35]

By now a general consensus emerged of a d-wave pairing state in the cuprates, with the superconducting gap changing sign around the nodal points along the [110]-direction on the Fermi surface. A simple inspection of the BCS gap equation shows that such a pairing state cannot be mediated by isotropic electron-phonon coupling. In fact, isotropic electron-phonon coupling would suppress a d-wave pairing state caused by another mechanism (e.g. spin fluctuations), because in such a case phonons would scatter quasiparticles between the parts of the Fermi surface with opposite sign of the superconducting gap.

A.S. Alexandrov showed that a d-wave state can be mediated by acoustic phonons with highly anisotropic electron-phonon coupling. [59] No experimental evidence has appeared up to now of any electron-phonon coupling for acoustic phonons, but the self-energy of the bond-stretching modes IS large and highly anisotropic (see fig. 20). In fact the bond-stretching phonons that couple most strongly to electrons connect the parts of the Fermi surface where the gap has the same sign, so these phonons will at least not interfere with d-wave superconductivity and may actually enhance $T_c$.

Strong renormalization of the bond stretching phonons has been taken as evidence for a soft collective charge mode [60,61] or an incipient instability [62] with respect to the formation of either polarons, biporarons [63,64,65], charge density wave order [66], phase separation [67,68,69], valence bond order [70], or other inhomogeneity [71]. These may or may not be related to the mechanism of stripe formation. A number of studies suggested that these instabilities may lead to superconductivity. [62,63,64,68]

S. Ishihara and N. Nagaosa examined the interplay of the interaction of the bond-stretching phonons with electrons in the t-J model and, after including vertex corrections, obtained the effective d-wave pairing interaction, F(qW), mediated by the bond-stretching phonons. [72] (see Fig. 21) In the absence of strong nesting as in the case of $La_{1.85}Sr_{0.15}CuO_4$, such a pairing interaction scales with the phonon-self-energy. Interestingly, the functional form of F(qW) that they obtained is similar to

Fig. 15, which shows the color map of anomalous phonon broadening. This result was a prediction, since it was obtained before the full picture of the bond-stretching phonon renormalization was known.

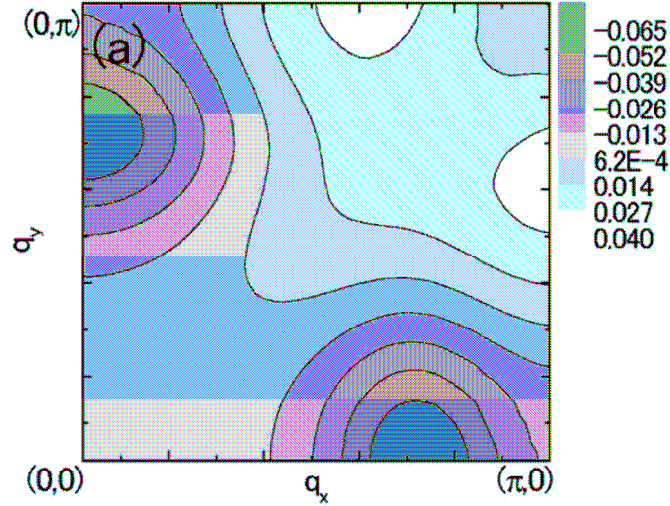

Figure 21. Contour map of the effective paring interaction F(qW) calculated from the t-J model with the off-diagonal coupling case and the vertex correction. (from Ref. [72])

They also showed that this interaction can mediate d-wave pairing.

## 3. Other Cuprates
### 3.1 YBa$_2$Cu$_3$O$_{6+x}$

It is necessary to establish the universality of the phonon anomalies observed in the La$_{2-x}$Sr$_x$CuO$_4$ family. Until now much less work has been performed on other cuprates, because they are more difficult to measure either due to higher background, no availability of large samples for INS, or low IXS scattering cross section.

In the case of YBa$_2$Cu$_3$O$_{6+x}$ the orthorhombic structure combined with twinning complicates the interpretation of the results. Very little work has been done so far on detwinned samples [73] because they are smaller than the twinned ones. Furthermore, two CuO$_2$ layers in the unit cell introduce two bond-stretching branches, of Δ1 and Δ4 symmetry.

At optimal doping bond-stretching phonons propagating along the chain direction show an anomaly that is in many respects similar to the one in La$_{2-x}$Sr$_x$CuO$_4$. [74,75] It is absent at 300K, and appears at low temperatures. Chung et al. [75] reported that the spectral weight of the bond-stretching phonons in the Δ1 symmetry redistributes to lower energies below the superconducting transition temperature, T$_c$=93K. (see Fig. 22)

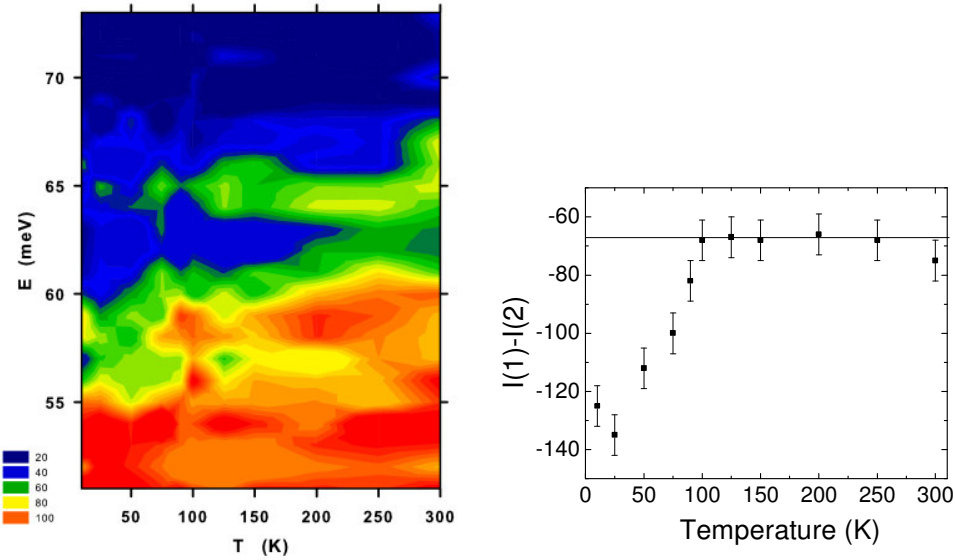

Figure 22. (right) Inelastic scattering intensity of YBa$_2$Cu$_3$O$_{6.95}$ at Q=(3.25,0,0) as a function of temperature, determined with the triple-axis spectrometer at the HFIR. Data were smoothed once to reduce noise. (left) Temperature dependence of the intensity difference (I1)-(I2), where I1 is the average intensity from 56 to 68 meV, I2 from 51 to 55 eV, at **Q**=(3.25,0,0). T$_c$ of the sample was 93K. (from Ref. [75])

L. Pintschovius et al. [74] and D. Reznik et al. [76] found that a similar transfer of spectral weight occurs for the Δ4 phonons but starting close to 200K, not at T$_c$. They interpreted this transfer of spectral weight as arising from softening of the bond-stretching phonon polarized along b*, which transfers its eigenvector to the branches that are lower in energy. This interpretation could explain the observed behavior with some important caveats, but more work is necessary to better understand this effect. Figure 23 shows that this transfer of spectral weight accelerated below T$_c$ saturating near 50K. While clearly related to the onset of superconductivity, this effect is not understood.

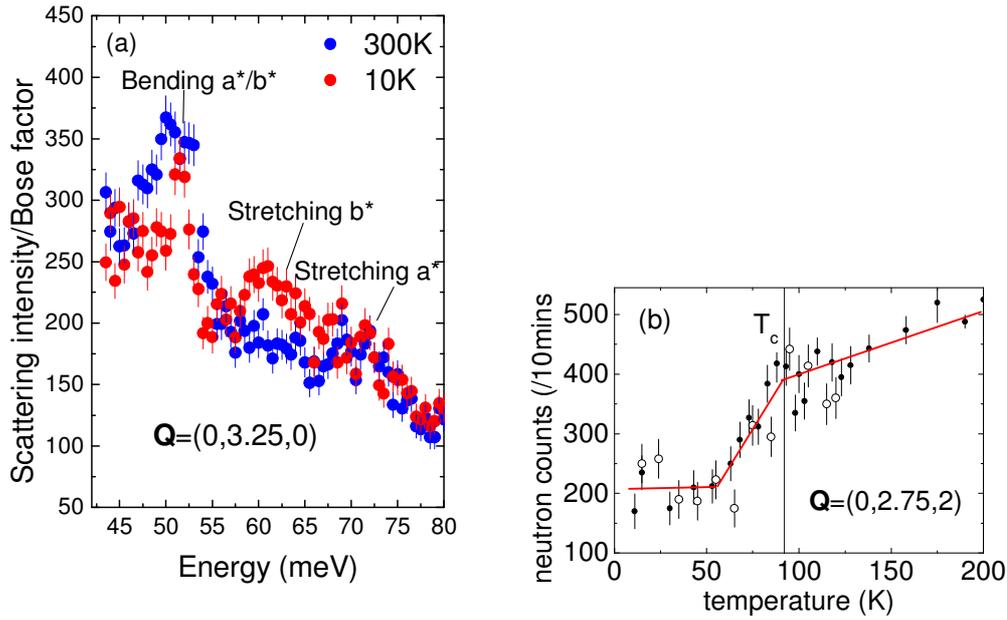

Figure 23. Data of Ref. [76] for the Δ4 symmetry in YBa$_2$Cu$_3$O$_{6.95}$. (a) Comparion of the 300K and 10K spectra. (b) Backround-subtracted intensity at Q = (0, 2.75,−2) and E = 60 meV (see text). Open and solid circles represent different datasets. From Ref. [76]

D. Reznik et al. also showed that the transfer of spectral weight in the Δ1 symmatry also begins well above T$_c$ with the most pronounced change below T$_c$. This result seems to contradict the observation of Chung et al. [75] who reported that the phonon intensity shift in Δ1 symmetry occurs only below T$_c$. According to Reznik et al. the effect would also appear only below T$_c$ if they excluded the intensity below 50 meV from their analysis [76] as was done in Ref. [75]. So in this respect the two studies are consistent.

The phonon anomaly in YBa$_2$Cu$_3$O$_{6.95}$ seems to extend far in the transverse direction (Fig. 25), i.e. it may be consistent with the 1D picture. [76] (also see sec. 2.5) However, twinning of the sample made the data difficult to interpret. Otherwise, the phonon anomaly in optimally-doped YBa$_2$Cu$_3$O$_{6.95}$ is similar to the effect in La$_{2-x}$Sr$_x$CuO$_4$. [1]

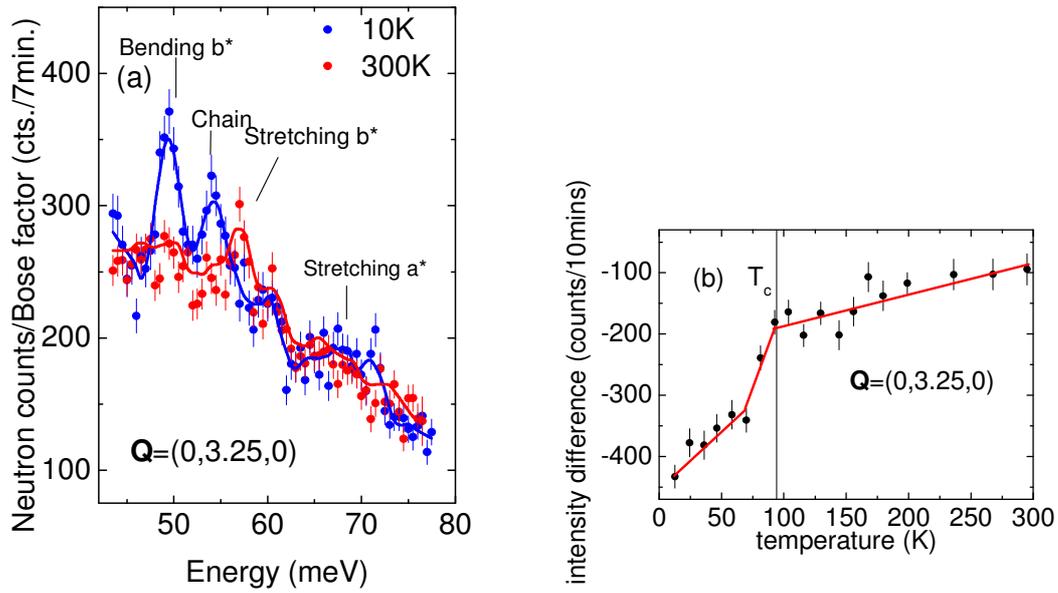

Figure 24. Data of Ref. [76] for the Δ1 symmetry. (a) Phonon spectra at 300K and 10K. The main difference with Ref. [75] and Fig. 22 is a bigger energy range here: 42-75meV in [76] vs. 51-72 meV in [75] (b) The difference between the intensity at 57 meV and the average of intensities at 53 meV and 49 meV at **Q** = (0, 3.25, 0). Temperature dependence above $T_c$ not seen in Fig. 22 comes from inclusion of the 49meV phonon, which falls outside the energy range investigated in [75].

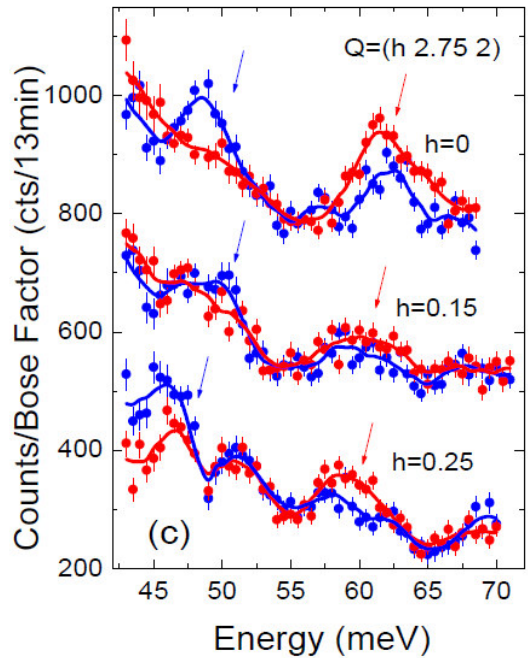

Figure 25. Energy (E) scans taken at 200K (red) and 10K (blue). Data were taken with the final energy $E_f$ = 13.4 meV for E > 46 meV and with $E_f$ = 12.5 meV for 43 <E <48 meV. The 12.5-meV data were corrected for the different resolution volume by multiplying by $(13.4/12.5)^2$. The resulting intensities were averaged in the overlapping energy range (46.5-48meV). The 200K data were divided by the Bose factor and 23 counts were subtracted to correct for the temperature dependence of the background. Blue/red arrows indicate intensity gain/loss. (from Ref. [76])

Much less is known about $YBa_2Cu_3O_{6+x}$ at lower doping levels. Stretzel et al [77] reported splitting of the bond-streching branch arguing in favor of charge inhomogeneity, whereas Pintschovius et al. [78] explained similar results in terms of the difference between the dispersion of the stretching phonons propagating parallel and perpendicular to Cu-O chains. This disagreement needs to be settled by measurements on detwinned samples.

### 4.2 $HgBa_2CuO_{4+x}$

Bond-stretching phonons in $HgBa_2CuO_{4+x}$ have been measured by Uchiyama et al. [79] These measurements showed that the bond-stretching phonons soften similarly to $La_{2-x}Sr_xCuO_4$ and $YBa_2Cu_3O_{6+x}$. It is important to extend this study to different dopings, temperatures and nonzero transverse wavevectors.

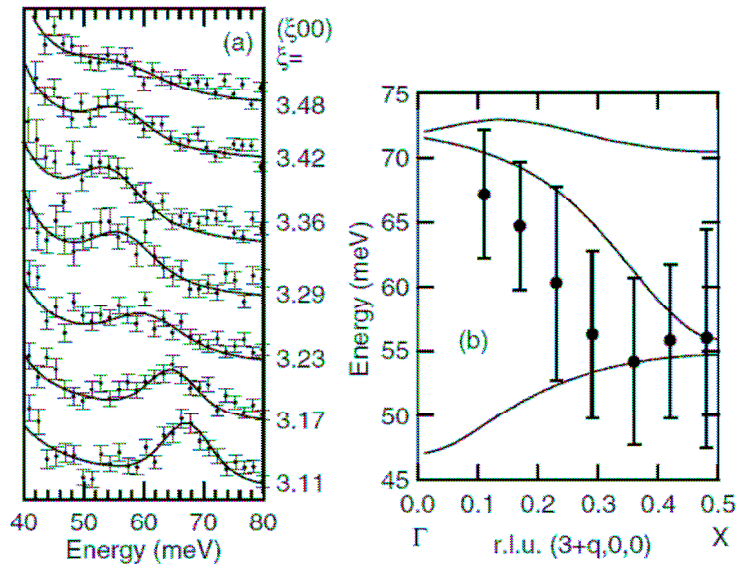

Figure 26. Bond stretching phonons in $HgBa_2CuO_{4+x}$. (from Ref. [79]) (a) Enlarged spectra taken close to the bond stretching mode plotted on a linear scale. (b) Data points represent frequencies of the bond-stretching phonons. The lines show the shell model calculation in which the interaction between the next-nearest neighbor oxygens in the $CuO_2$ plane is added. The lines indicate (top to bottom) the *c*-polarized apical oxygen mode, the *a*-polarized Cu-O bond stretching mode, and the *a*-polarized in-plane Cu-O bending mode, respectively. The vertical bars indicate the FWHM of the peaks determined in fitting data shown in (a).

### 4.3 $Bi_2Sr_{1.6}La_{0.4}Cu_2O_{6+x}$

Graf et al. [80] measured phonon dispersions by IXS and electronic dispersions by ARPES in a single-layer Bi-based cuprate, $Bi_2Sr_{1.6}La_{0.4}Cu_2O_{6+x}$. They reported a similar phonon anomaly as in other cuprates and argued in favor of a correlation between this phonon anomaly, the kink observed in photoemission and the Fermi arc that characterizes the pseudogap phase. They related the sudden onset of phonon broadening near **q**=(0.2,0,0) to coupling of the phonon to the Fermi arc region of the Fermi surface, but not to the pseudogap region. The Fermi arc region is not nested, so exceptionally large electron-phonon coupling for the stretching branch is necessary for this interpretation to be valid. (as in Cr as described in Sec. 2) In addition one needs to consider that in other families of cuprates, where the doping dependence has been investigated, the wavevector of the onset of the phonon effect is doping independent, whereas the length of the Fermi arc strongly depends on doping. More detailed studies of this compound, especially as a function of doping, are necessary to clarify these issues.

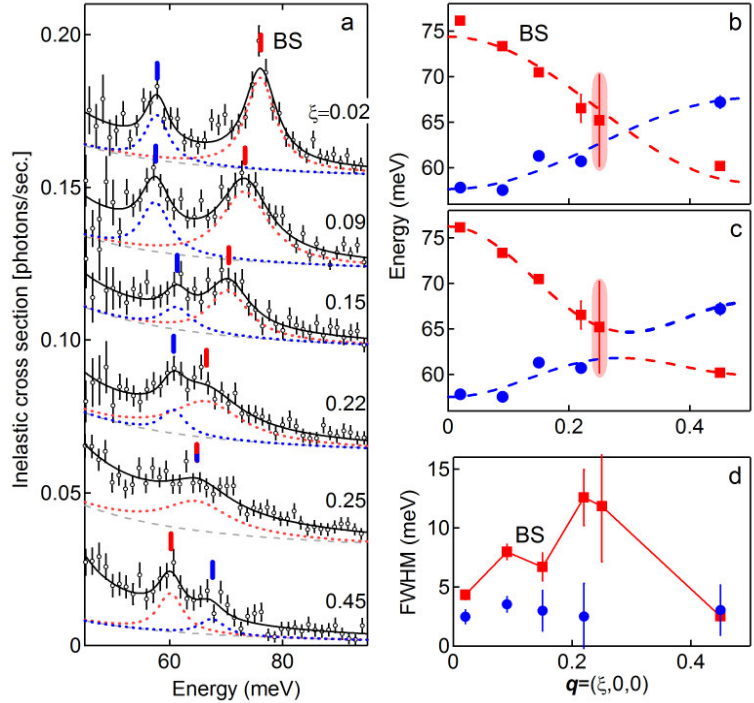

Figure 27. LO phonon dispersions in $Bi_2Sr_{1.6}La_{0.4}Cu_2O_{6+x}$. [80] (a) IXS spectra for $Q=(3+\xi,0,0)$ with $\xi$ from the BZ center (top spectrum, $\xi = 0.02$) to the BZ boundary (bottom spectrum, $\xi = 0.45$). The spectra are vertically shifted. The solid lines show the harmonic oscillator fit, the dashed lines show the elastic tail and the dotted lines show the two modes used in the fit. (b,c) Phonon dispersions and linewidths. The cosine dashed lines are guides for the eyes illustrating the crossing (b) and anticrossing (c) scenarios. (d) Full width at half maximum. The error bars are an estimate of the standard deviation of the fit coefficients

**4.4 Electron-doped cuprates**

Bond-stretching phonons have been investigated in electron-doped cuprates only in $Nd_{2-x}Ce_xCuO_4$. Phonon density of states measurements on powder samples showed that electron doping softens the highest energy oxygen phonons as occurs in the case of hole-doping. [81] The first single crystal experiment has been performed by d'Astuto et al. by IXS [82] who found that the bond-stretching phonon branch dispersed steeply downwards beyond h=0.15. This work was, in fact, the first IXS experiment on the high $T_c$ cuprates. These measurements, however were complicated by the anticrossing of the bond-stretching branch with another branch due to Nd-O vibrations that dispersed sharply upwards. The anticrossing occurs near h=0.2 complicating the interpretation of the data near these wavevectors. Another difficulty came from low IXS scattering cross sections for the oxygen vibrations.

A neutron scattering investigation has been performed by M. Braden et al. [83] once large single crystals became available. Oxygen phonons have a higher scattering cross section in the INS than in the IXS experiments, allowing a more accurate determination of the phonon dispersions.

The two studies showed that most the bond-stretching phonon dispersion in $Nd_{1.85}Ce_{0.15}CuO_4$ was similar to that in the hole-doped compounds. (Fig. 28) This similarity points at a commonality between the tendencies to charge inhomogeneity between the hole-doped and electron-doped compounds as discussed in Ref. [83].

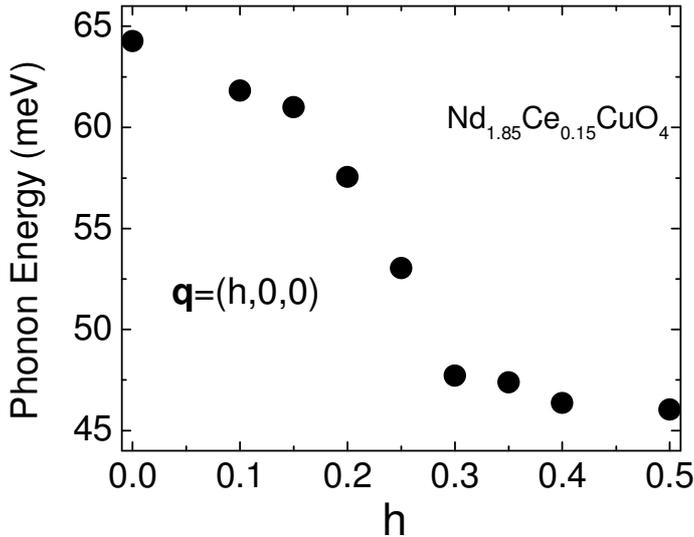

Figure 28. Dispersion of the Cu-O bond-stretching phonon in $Nd_{1.85}Ce_{0.15}CuO_4$ adapted from Ref. [83]

## 5. Conclusions

I hope to have shown that a lot of progress has been made in recent years in understanding the phenomenology of the giant electron-phonon coupling of the bond-stretching phonons in $La_{2-x}Sr_xCuO_4$. The picture that emerged is that the bond-stretching phonons around **q**=(0.3,0,0) are softer and broader than expected from conventional theory. This effect may be related to incipient instability with respect to the formation of dynamic stripes or another charge-ordered or inhomogeneous state. Doping dependence of this phonon anomaly suggests that it is associated with the mechanism of superconductivity.

Much more experimental and theoretical work is necessary to understand the role of these phonons in superconductivity. The most important shortcoming of the present understanding is that $La_{2-x}Sr_xCuO_4$ has a relatively low $T_c$ and a number of its properties (for example the magnetic spectrum) differ in important ways from the cuprates with higher $T_c$s. Thus it is essential to put more emphasis on investigating cuprates with higher transition temperatures where similar phonon anomalies have been already found, but many open questions still remain.

## 6. Acknowledgement

I greatly benefited from interactions with many people over the years without which this work would not have been possible. In particular, I would like to acknowledge discussions with L. Pintschovius, R. Heid, K.-P. Bohnen, W. Reichardt, H. v. Löhneysen, F. Weber, D. Lamago, A. Hamann, J.M. Tranquada, S.A. Kivelson, T. Egami, Y. Endoh, M. Arai, K. Yamada, P.B. Allen, I.I. Mazin, J. Zaanen, D.J. Singh, G. Khaliullin, B. Keimer, D.A. Neumann, J.W. Lynn, A. Mischenko, N. Nagaosa, F. Onufrieva, P. Pfeuty, P. Bourges, Y. Sidis, O. Gunnarson, S.I. Mukhin, P. Horsch, T.P. Devereaux, Z.-X. Shen, A.Q.R. Baron, D.S. Dessau, P. Böni, M. d'Astuto, A. Lanzara, M. Greven, F. Giustino, J. Noffsinger, and M. Hoesch.